\documentstyle[preprint,prl,aps]{revtex}
\begin{document}
\draft
\title{ Species Doubling and Chiral Lagrangians }
\author{Michael Creutz and Michel Tytgat
\thanks{This manuscript has been authored under contract number
DE-AC02-76CH00016 with the U.S.~Department of Energy.  Accordingly,
the U.S.~Government retains a non-exclusive, royalty-free license to
publish or reproduce the published form of this contribution, or allow
others to do so, for U.S.~Government purposes.}}

\address{Physics Department,
Brookhaven National Laboratory,
Upton, NY 11973; email:
creutz@bnl.gov}

\date{April, 1996}
\maketitle

\begin{abstract}
Coupling gauge fields to the chiral currents from an effective
Lagrangian for pseudoscalar mesons naturally gives rise to a species
doubling phenomenon similar to that seen with fermionic fields in
lattice gauge theory.
\end {abstract}


Species doubling is one of the oldest puzzles in lattice gauge theory.
Naive fermion formulations on a lattice are plagued by the appearance
of spurious low energy states.  Various schemes have been implemented
to remove the extra particles, but usually at the expense of
mutilating valid symmetries.  Only recently has a lattice formulation
been presented that elegantly preserves the underlying chiral
invariance of the strong interactions [1].

The problem is intricately entwined with the famous axial anomalies.
A lattice regulator removes all infinities; so, anomalies require an
explicit symmetry breaking at the outset.  This is also familiar from
the Pauli-Villars [2] approach, where the mass of the heavy regulator
field is not chirally symmetric.  As the regulator is removed, it
leaves a remnant determining an overall chiral phase.  When the
physical quarks maintain a finite mass, this phase is observable as
the well known strong CP parameter $\theta$ [3].

Going beyond purely hadronic physics to the gauge interactions of the
electroweak theory, non-perturbative chiral issues remain unresolved.
The $W$ bosons couple in an inherently parity violating manner, and
rely for consistency on a subtle anomaly cancellation between quark and
lepton contributions.  While a flurry of recent work has advocated
treating the fermions with a separate limit [4-6], it remains unknown
how to implement this cancellation in a fully finite and
gauge-invariant lattice theory.  It is tempting to speculate that
there is a shortcoming in either the lattice approach or in the
standard model.

Here we argue that the so called ``doubling'' problem is not unique to
the lattice approach, but is a more general consequence of chiral
anomalies.  Starting with a effective Lagrangian for the strong
interactions of the pseudoscalar mesons, we consider coupling gauge
fields to the $SU(n_f)\times SU(n_f)$ symmetries of this model.  When
these gauge fields are themselves chiral, the gauging process
naturally introduces additional Goldstone fields mirroring the
original theory.  As with the lattice doublers, the mirror fields
cancel anomalies.  The issue reduces to non-perturbatively removing
the extra species when the original theory is anomaly free.

In the chiral Lagrangian approach, the effects of anomalies are
summarized in a term discussed some time ago by Wess and Zumino [7],
and later elucidated by Witten [8].  This coupling requires extending
the fields into an internal space, only the boundary of which is
relevant to the equations of motion.  On adding a coupling to a local
gauge field, however, the boundary can acquire additional
contributions.  The essence of this paper is that these are most
naturally written in terms of doubler fields.

Ref.~[9] treats a canonical quantization of anomalous fermion
theories.  For consistency they couple the gauge fields to a
Wess-Zumino term.  While their starting point was rather different,
they reach a similar conclusion that anomalous theories naturally lead
to the introduction of new degrees of freedom.

To start, we briefly review the basic philosophy behind the effective
Lagrangian approach.  Our underlying theory contains a set of
fermionic quark fields $\psi^a(x)$ interacting with non-Abelian gauge
fields.  Here we suppress all indices except flavor, represented by the
index $a$, and space-time, represented by $x$.  From $\psi$ we project
out right and left handed parts, $\psi^a_R={1\over 2}
(1+\gamma_5)\psi^a$ and $\psi^a_L={1\over 2} (1-\gamma_5)\psi^a$.  For
the purpose of this discussion we ignore fermion masses; such could be
introduced, as in the standard model, via a Higgs mechanism.

The underlying quark-gluon theory with massless quarks is invariant
under a global $SU(n_f)\times SU(n_f)$ symmetry, where $n_f$
represents the number of flavors.  Under this, the quark fields
transform as $\psi_L^a\rightarrow \psi^b g_L^{ba}$ and
$\psi_R^a\rightarrow \psi^b g_R^{ba}$.  Here $g_L$ and $g_R$ are
elements of $SU(n_f)$.  Formally the classical Lagrangian is also
invariant under a global $U(1)\times U(1)$ symmetry of phases for the
left and right quark fields, but the axial part of the latter symmetry
is broken by quantum effects, leaving just the vector $U(1)$ of
fermion number.  While anomalies also play the key role in this
breaking, that is not the subject of this paper.

In the conventional view, the axial part of the global chiral symmetry
is spontaneously broken by the vacuum, resulting in $n_f^2-1$
Goldstone bosons and a remaining explicit $SU(n_f)$ flavor symmetry.
This is usually described via the composite field $\overline \psi^a_R
\psi^b_L$ acquiring a vacuum expectation value.  Without flavor
breaking, one can use the chiral symmetry to pick a standard vacuum
with, say, $\langle\overline \psi^a_R \psi^b_L\rangle=v\delta^{ab}$.
Here the parameter $v$ determining the magnitude of the expectation
value requires a renormalization scheme for precise definition.  The
vacuum is degenerate (after the usual extension of the quantum Hilbert
space to a Banach space), and one could choose to replace
$\delta^{ab}$ by an arbitrary element $g^{ab}$ of $SU(n_f)$.  The
basic idea of the effective Lagrangian is to promote this element into
a local field $g(x)$.  Slow variations of this field represent the
Goldstone bosons arising from the degeneracy of the vacuum.

Equivalently, imagine integrating out the fermionic fields under a
constraint $\langle\overline \psi^a_R \psi^b_L\rangle=v g^{ab}(x)$,
ignore the massive modes associated with fluctuations in $v$, and use
the resulting path integral to define an effective theory for $g(x)\in
SU(n_f)$.  The chiral symmetry becomes an invariance under
$g(x)\rightarrow g_L^\dagger g(x) g_R$ for arbitrary $g_L$ and $g_R$.

More quantitatively, the approach represents an expansion in powers of
the momenta of the light particles [10].  The lowest order action
contains the first term in this expansion
$$
S_0={F_\pi^2\over 4}\int d^4x\ {\rm Tr}(\partial_\mu g\partial_\mu g^\dagger). 
\eqno (1)
$$
The numerical constant $F_\pi$ sets the scale and has an experimental
value around 93 MeV.  To relate this to conventionally normalized pion
fields, we define $g=\exp(i\pi\cdot\lambda/F_\pi)$ where the
$n_f^2-1$ matrices $\lambda$ generate $SU(n_f)$ and are normalized ${\rm Tr}
\lambda^\alpha \lambda^\beta=2\delta^{\alpha\beta}$.

From this lowest order action, the equations of motion are $\partial_\mu
J_{L,\mu}^\alpha=0$, where the ``left'' current is 
$$
J_{L,\mu}^\alpha= {iF_\pi^2\over 4}{\rm Tr}\lambda^\alpha(\partial_\mu g)g^\dagger
\eqno (2)
$$
These equations have an equivalent form involving ``right'' currents
$J_{R,\mu}^\alpha={iF_\pi^2\over 4}{\rm Tr}\lambda^\alpha
g^\dagger\partial_\mu g$.  There is a vast literature about adding
higher derivative terms the above action [10].  This, however, is not
what this paper is about.

We are interested in a special higher derivative coupling which is
necessarily present and describes the effects of anomalies from the
underlying quark fields.  As is well known, this term is curious in
that it cannot be written simply as an integral of a local expression
in $g(x)$, even though the resulting contribution to the equations of
motion is fully local [7,8,11].  Continuing to write the equations of
motion in terms of a current, a possible addition which satisfies the
required symmetries is
$$
J_{L,\mu}^\alpha=
{iF_\pi^2\over 4}{\rm Tr}\lambda^\alpha(\partial_\mu g)g^\dagger
+{in_c\over 48\pi^2} \epsilon_{\mu\nu\rho\sigma}{\rm Tr}\lambda^\alpha
(\partial_\nu g)g^\dagger
(\partial_\rho g)g^\dagger
(\partial_\sigma g)g^\dagger
\eqno (3)
$$
The equations of motion remain that the current be divergence less,
$\partial_\mu J_{L,\mu}^\alpha=0$.

The addition in Eq.~(3) is the simplest possible term involving the
antisymmetric tensor, a shadow of the factors of $\gamma_5$ involved
in the chiral anomalies.  As is also well known [8], quantum mechanics
requires the dimensionless coupling strength $n_c$ to be an integer
corresponding to the number of degrees of freedom (``colors'') in the
underlying confining theory.  Thus this term must indeed be present.

Continuing with this lightning review, we desire an action which
generates the above equations of motion.  This requires extending the
field $g(x)$ beyond a simple mapping of space-time into the group.
For this purpose, we introduce an auxiliary variable $s$ to interpolate
between the field $g(x)$ and some fixed group element $g_0$.  Thus we
consider an extended field $h(x,s)$ satisfying $h(x,1)=g(x)$ and
$h(x,0)=g_0$.  This extension is not unique, but the final equations
of motion are independent of the chosen path.  We now write the action
$$
S={F_\pi^2\over 4}\int d^4x\ 
{\rm Tr}(\partial_\mu g\partial_\mu g^\dagger)
+{n_c\over 240 \pi^2}\int d^4x\int_0^1 
ds\ \epsilon_{\alpha\beta\gamma\delta\rho}
{\rm Tr} h_\alpha h_\beta h_\gamma h_\delta h_\rho.
\eqno (4)
$$
Here we introduce the shorthand notation $h_\alpha=i(\partial_\alpha h)
h^\dagger$ and regard $s$ as a fifth coordinate.  The antisymmetric
tensor satisfies $\epsilon_{1,2,3,4,5}=1$.

To find the equations of motion, consider a small variation of
$h(x,s)$.  This can be shown to change the final integrand by a total
divergence, which then integrates to a surface term.  Working with
either spherical or toroidal boundary conditions in the space-time
directions, this surface only involves the boundaries of the $s$
integration.  When $s=0$, space-time derivatives acting on the
constant matrix $g_0$ will vanish.  The surface at $s=1$ generates
precisely the desired additional term in Eq.~(3).

Geometrically, the last term in Eq.~(4) is the volume of a piece of
the $S_5$ sphere appearing in the structure of $SU(n_f)$ for $n_f\ge
3$.  The mapping of four dimensional space-time into the group
surrounds this volume.  Global chiral rotations merely shift this
region around, leaving its numerical volume invariant.  As emphasized
by Witten [8], this volume is only defined up to a multiple of the
total volume of the $S_5$ mapping into the gauge group.  Different
extensions into the $s$ coordinate can modify the above five
dimensional integral by an integer multiple of $480\pi^3$.  To have a
well defined quantum theory, the action must be determined up to a
multiple of $2\pi$.  Thus the quantization of $n_c$ to an integer
value, much like the charge of a magnetic monopole.

Crucial to this discussion is the irrelevance of the starting group
element $g_0$ and the lower end of the $s$ integration.  The main
point of this paper is to emphasize the difficulty of maintaining this
condition when the symmetries become local.  In particular, we want to
extend the symmetry and allow $g_{R,L}$ to depend on the space-time
coordinate $x$.  As usual, this requires the introduction of local
gauge fields.  When we make the transformation $g(x)\rightarrow
g_L^\dagger(x) g(x) g_R(x)$, derivatives of $g$ transform as
$$
\partial_\mu g\longrightarrow
g_L^\dagger\left (
\partial_\mu g-\partial_\mu g_L g_L^\dagger g+ g\partial_\mu g_R g_R^\dagger 
\right) g_R
\eqno (5)
$$
To compensate, we introduce left and right gauge fields transforming as
$$\matrix{
A_{L,\mu} & \longrightarrow
& g_L^\dagger A_{L,\mu} g_L + i g_L^\dagger\partial_\mu g_L \cr
A_{R,\mu} & \longrightarrow 
& g_R^\dagger A_{R,\mu} g_R + i g_R^\dagger\partial_\mu g_R \cr
}
\eqno (6)
$$
Then the combination
$$
D_\mu g = \partial_\mu g-iA_{L,\mu} g + ig A_{R,\mu}
\eqno (7)
$$ 
transforms nicely: $D_\mu g\rightarrow g_L^\dagger D_\mu g g_R$.  If we
make the generalized minimal replacement $\partial_\mu g\rightarrow
D_\mu g$ in $S_0$, we find a gauge invariant action.

A problem arises when we go on to the Wess-Zumino term.  We require a
prescription for the gauge transformation on the interpolated group
element $h(x,s)$.  For this purpose, note a striking analogy with the
domain wall approach to chiral fermions first promoted by Kaplan[12].
There an extra dimension was also introduced, with the fermions being
surface modes bound to a four dimensional interface.  The usual
approach to adding gauge fields involves, first, not giving the gauge
fields a dependence on the extra coordinate, and second, forcing the
component of the gauge field pointing in the extra dimension to vanish
[13-14].  To be more precise, in terms of a five dimensional gauge
field, we take $A_\mu(x,s)=A_\mu(x)$ and $A_s=0$ for both the left and
right handed parts.  Relaxing either of these would introduce extra
degrees of freedom for which there is no desire.  Thus the natural
extension of the gauge transformation to all values of $s$ is to take
$h(x,s)\rightarrow g_L^\dagger(x) h(x,s) g_R(x)$ with $g_{L,R}$
independent of $s$.

With this prescription for interpolating the gauge fields into the $s$
dimension, we replace the derivatives in the Wess-Zumino term with
covariant derivatives, similar to Eq.~(7).  This alone does not give
equations of motion independent of the interpolation into the extra
dimension.  However, adding terms linear and quadratic in the gauge
field strengths allows construction of a five dimensional Wess-Zumino
term for which variations are again a total derivative.  The term is
still ambiguous up to non-minimal couplings.  For the photon, parity
invariance uniquely fixes such terms, but this goes beyond the subject
of this paper.

This procedure works well for a vector-like gauge field, where we take
$g_L(x)=g_R(x)$ and $A_L=A_R$.  We could, for example, take $g_0$ to be
the identity, and then the gauge transformation cancels out at $s=0$.
The approach gives the coupling of the photon field to the
pseudoscalar mesons, including [8] a piece that describes
$\pi\rightarrow 2\gamma$.  This supports the necessity of the
Wess-Zumino term, and determines the coefficient to be proportional to
the dimension of the quark representation in the underlying confining
symmetry group.

A difficulty arises when coupling a gauge field to an axial current.
For example, the weak bosons of the standard electroweak theory
involve such a coupling.  In this case the above prescription at $s=0$
takes $g_0\rightarrow g_L^\dagger(x) g_0 g_R(x)$, which in general
will no longer be a constant group element.  After a gauge
transformation, variations of the action give new non-vanishing
contributions to the equations of motion from the lower end of the $s$
integration.

The simplest solution promotes the $s=0$ fields to be dynamical.  Thus
we replace the field $g(x)$ with two fields $g_0(x)$ and $g_1(x)$.  The
interpolating field now has the properties $h(x,0)=g_0(x)$ and
$h(x,1)=g_1(x)$.  The action becomes
$$
S={F_\pi^2\over 4}
\int d^4x\ {\rm Tr}(D_\mu g_0D_\mu g_0^\dagger+D_\mu g_1D_\mu g_1^\dagger) 
+ \Gamma
\eqno (8)
$$
where $\Gamma$ denotes the appropriatly gauged Wess-Zumino term.

While we now have a gauge invariant theory, it differs from the
starting theory through doubling of meson species.  The extra
particles are associated with the second set of group valued fields
$g_0(x)$.  The Wess-Zumino term of the new fields has the
opposite sign since it comes from the lower end of the $s$
integration.  Thus, these ``mirror'' particles have reflected chiral
properties and implement a cancellation of all anomalies.  In essence,
we have circumvented the subtleties in gauging the model.

The value of $F_\pi$ need not be the same for $g_0$ and $g_1$; so,
their strong interactions might differ in scale.  Nevertheless,
coupling with equal magnitude to the gauge bosons, the new fields
cannot be ignored.  The doubling is less severe than in the lattice
approach, where each pairing in the number of fermion fields gives a
factor of four in boson species.

Had we only coupled gauge fields to vector currents, we could easily
remove the doublers using a diagonal mass term at $s=0$.  For example,
with a term $M{\rm Tr} g_0(x)$ added to the Lagrangian density, $M$
could be arbitrarily large, forcing $g_0$ towards the identity.  Such
a term is invariant under vector rotations, but not under axial
symmetries.

The doublers arise in complete analogy to the problems appearing in
the surface mode approach to chiral lattice fermions [12-15].  In both
cases, an extension to an extra dimension is introduced.  Difficulties
arise from the appearance of an extra interface in the $s$ coordinate.
This new surface cannot be ignored since it couples with equal
strength to the gauge fields.

If we relax the constraints and let $g_{L,R}$ depend on $s$, we expect
problems similar to those seen with domain wall fermions.  In
particular, when the gauge fields vary in the extra dimension, four
dimensional gauge invariance is lost.  Symmetry can be restored via a
Higgs field, but this introduces the possibility of unwanted degrees
of freedom in the physical spectrum.  Ref.~[14] explores the
possibility of sharply truncating the gauge field at an intermediate
value of the extra coordinate.  This gives rise to new low energy
bound states acting much like the undesired doubler states.

Introducing a Higgs field does permit different masses for the extra
species.  In particular, the matter couplings of the Higgs field can
depend on $s$.  Qualitative arguments suggest that triviality effects
on such couplings limit their strength, precluding masses for the
extra species beyond a typical weak interaction scale.  Presumably
such constraints will be the strongest when the anomalies in the
undoubled sector are not properly canceled.  With domain-wall
fermions, taking the Higgs-fermion coupling to infinity on one wall
introduces a plethora of new low energy bound states [15].

These problems reemphasize the subtle way the standard model cancels
anomalies between the quarks and the leptons.  If the contributions of
the leptons are ignored, no non-perturbative approach can be expected
to accommodate gauged weak currents.  Indeed, the doubling discussed
here arises as a necessary consequence of residual anomalies.  When
the required cancellations occur between different fermion
representations, perturbation theory appears to be consistent, while
all known non-perturbative approaches remain awkward.  

There are several possible solutions to these doubling problems.
Least interesting would be some trivial missed issue in our search for
a non-perturbative definition of a chiral gauge theory.  On the other
hand, mirror particles might actually exist, perhaps with masses
comparable to the weak scale [16].  Such extra fields might even be
useful in the spontaneous breaking of the electroweak theory [17].  A
related alternative has the standard model arise from the spontaneous
breaking of an underlying vector-like unified theory containing
additional heavy bosons coupling with opposite parity fermions [18].
All of these involve a profusion of new particles awaiting discovery.
A speculative solution would twist the extra dimension so that the
doubling particles could be among those already observed.  This
requires the interpolation in the extra dimension to mix the quarks
and the leptons, all of which are involved in the anomaly
cancellations of the standard model.  While such a scheme remains
elusive, it conceivably could require the existence of multiple
families.

\section*{Acknowledgments} We are grateful to R. Pisarski and C. Rebbi
for stimulating discussions.

 \end{document}